# Temporal Logics on Words with Multiple Data Values[*]

Ahmet Kara, Thomas Schwentick, Thomas Zeume
TU Dortmund, Germany
{ahmet.kara, thomas.schwentick, thomas.zeume}@cs.tu-dortmund.de

Monday 15$^{\text{th}}$ October, 2018

**Abstract**

The paper proposes and studies temporal logics for attributed words, that is, data words with a (finite) set of (attribute,value)-pairs at each position. It considers a basic logic which is a semantical fragment of the logic LTL$_1^\downarrow$ of Demri and Lazic with operators for navigation into the future and the past. By reduction to the emptiness problem for data automata it is shown that this basic logic is decidable. Whereas the basic logic only allows navigation to positions where a fixed data value occurs, extensions are studied that also allow navigation to positions with different data values. Besides some undecidable results it is shown that the extension by a certain UNTIL-operator with an inequality target condition remains decidable.

## 1 Introduction

Motivated by questions from XML theory and automated verification, extensions of (finite or infinite) strings by data values from unbounded domains have been studied intensely in recent years. Various logics and automata for such data words have been invented and investigated.

A very early study by Kaminski and Francez [17] considered automata on strings over an "infinite alphabet". In [7], *data words* were invented as finite sequences of pairs $(\sigma, d)$, where $\sigma$ is a symbol from a finite alphabet and $d$ a value from a possibly infinite domain. In [6] multi-dimensional data words were considered where every position carries $N$ variable valuations, for some fixed $N$. Similar models can be found for instance in [2] and other work on parameterized verification. More powerful models were investigated in [19] and [14] where every position is labeled by the state of a relational database, i.e., by a set of relations over a fixed signature.

---

[*]We acknowledge the financial support by the German DFG under grant SCHW 678/4-1.

For the basic model of data strings with one data value per position a couple of automata models and logics have been invented and their algorithmic and expressive properties have been studied. On the automata side we mention register automata [17, 8, 22] (named *finite memory automata* in [17, 8]), pebble automata [22, 24], alternating 1-register automata [12], data automata [5] (or the equivalent class memory automata [3]).

On the logical side, classical logics like two-variable first-order logic [5] have been studied and recently order comparisons between data values have been considered [20, 23]. The satisfiability problem for two-variable first-order logic over data words is decidable if data values can only be compared for equality but positions can be compared with respect to the linear order and the successor relation [5]. However, the complexity is unknown. It is elementary if and only if testing reachability in Petri nets is elementary as well [5]. The proof of decidability uses *data automata*, a strong automata model with decidable non-emptiness.

More relevant for this paper are previous investigations of temporal logics on data words. A pioneering contribution was by Demri and Lazic [12] (the journal version of [11]) which introduced Freeze LTL. In a nutshell, Freeze LTL extends LTL by *freeze quantifiers* which[1] allow to "store" the current data value in a register and to test at a possibly different position whether that position carries the same value. Freeze LTL has a decidable finite satisfiability problem if it is restricted to one register ($LTL_1^\downarrow$) and to future navigation, but the complexity is not primitive recursive. With one register and past (and future) navigation it is undecidable. In [16] it is shown that these lower bounds even hold if only navigation with F and P (but without X) are allowed.

In [12], also a restriction of $LTL_1^\downarrow$, *simple $LTL_1^\downarrow$*, was investigated and it was shown that it is expressively equivalent to two-variable logics. The restriction requires that (syntactically) between each value test and the corresponding freeze quantifier there is at most one temporal operator and it disallows Until and Since navigation but allows past navigation. Thanks to the (effective) equivalence to two-variable logics, simple $LTL_1^\downarrow$ is decidable.

One of our aims in this paper was to find a decidable temporal logic on data words with past navigation that is more expressive than simple $LTL_1^\downarrow$. In particular it should allow Until navigation with reference to data values. On the other hand, the logics we study are semantical fragments of $LTL_1^\downarrow$. Furthermore this work was motivated by the decidable logic $CLTL^\diamond$ for multi-attribute data words [10]. It allows to test whether somewhere in the future (or past) a current data value occurs and it can compare data values between two positions of bounded distance. The logics proposed in this paper are intended to have more expressive power than $CLTL^\diamond$ while retaining its decidability.

---

[1]We note that the freeze quantifier itself was used already in [9] and in previous work, e.g., in [1].

**Contribution**

We propose and investigate temporal logics for multi-attribute data words. An attributed word is a string which can have a finite number of (attribute,value)-pairs at each position (in the spirit of XML) and has propositions rather than symbols (in the spirit of LTL).

We first define Basic Data LTL which mimics the navigation abilities of simple $LTL_1^{\downarrow}$, if only positive register tests are used. As sequences of such navigation steps do not do any harm we drop the requirement to freeze the data value at every step and replace freeze quantifiers by a class quantifier which restricts a sub-formula to the positions at which this data value appears. We show that a slight extension of this logic captures simple $LTL_1^{\downarrow}$ (Proposition 2) and that it is decidable (Theorem 2). Although strictly more expressive than $CLTL^{\diamond}$, the decidability proof for Basic Data LTL is conceptually simpler than the proof given in [10]. It uses an encoding of multi-attribute words by data words and a reduction to non-emptiness of data automata. A similar multi-attribute encoding has already been used in [13]. The result generalizes to attributed $\omega$-words (Theorem 3). Some obvious extensions (by navigation with respect to two data values or Until navigation where intermediate positions can be tested by data-free formulas) are undecidable (Theorems 4 and 6, respectively).

Finally, we add a powerful Until-operator to Basic Data LTL, which allows to navigate to a position with a data value that is *different* from the value of a given attribute at the starting position. Furthermore, it can test properties of intermediate positions by arbitrary sub-formulas and can even test (in a limited way) whether intermediate positions have attribute values different from or equal to the value on the starting position. The resulting logic can express all properties expessible in two-variable first-order logic and contains the Until operator. That this logic is still decidable is the main technical contribution of the paper.

The paper is organized as follows. In Section 2, we define attributed words and Basic Data LTL and give some example properties. In Section 3, we compare Basic Data LTL with other logics. Section 4 shows that Basic Data LTL is decidable and presents undecidability results for some extensions. Section 5 introduces the extended Until operator and shows decidability of the resulting logic. It also shows (the simple fact) that an Until-operator that navigates with respect to equality and allows (only) data-free intermediate tests quickly leads to an undecidable logic. We conclude in Section 6.

**Related work**

We discussed many related papers above. Another approach, combining temporal and classical logics, was studied in [14]. It allows to navigate by temporal operators and to evaluate first-order formulas in states. Properties depending on values at different states can be stated by global universal quantification of values. In [6] a first-order logic on multi-dimensional data words was studied.

**Acknowledgements**

The idea to extend the temporal logic that is equivalent to two-variable logics by Until operators (without reference to data) goes back to a suggestion by Mikołaj Bojanczyk [4]. We are also indebted to Volker Weber with whom we carried out first investigations before he tragically passed away in 2009. The remarks by the reviewers of FSTTCS 2010 helped to improve the presentation and to add some additional references.

## 2  Definitions

We first fix the data model and define BD-LTL afterwards. Finally we give an example that illustrates the way in which properties can be expressed

### 2.1  Attributed words

Let $\mathcal{PROP}$ and $\mathcal{ATT}$ be (possibly infinite) sets of propositions and attributes and $\mathcal{D}$ an infinite set of data values. An *attributed word* $w$ is a finite word where every position carries a finite set $\{p_1, \ldots, p_l\}$ of propositions from $\mathcal{PROP}$ and a finite set $\{(a_1, d_1), \ldots, (a_k, d_k) \mid a_i \neq a_j \text{ for } i \neq j\}$ of attribute-value pairs from $\mathcal{ATT} \times \mathcal{D}$.

Given an attributed word $w$ we denote the proposition set of position $i$ in $w$ by $w[i].\mathcal{P}$. A position $i$ is a *p-position* if $p \in w[i].\mathcal{P}$. By $w[i].@a$ we denote the value of attribute $a$ on position $i$. If position $i$ does not carry attribute $a$, then $w[i].@a = nil \notin D$. The *word projection* of an attributed word $w = w_1 \ldots w_n$ is defined by $str(w) := w[1].\mathcal{P} \ldots w[n].\mathcal{P}$. By $pos_d(w)$ we denote the set of *class positions* of $d$ in $w$, that is, the set of positions of $w$ with at least one attribute with value $d$. The *class word* $class_d(w)$ of $w$ with respect to $d$ is the restriction of $w$ to the positions of $pos_d(w)$.

We always consider sets of words over some finite set $\mathcal{P}$ of propositions and a finite set $\mathcal{V}$ of attributes[2]. We call an attributed word $w$ $\mathcal{V}$-*complete* for a finite set $\mathcal{V} \subseteq \mathcal{ATT}$ if every position of $w$ has exactly one pair $(a, d_a)$ for each $a \in \mathcal{V}$. A $\{a\}$-complete word is called *1-attributed word*. We refer to the value of attribute $@a$ at a position $i$ in a 1-attributed word as *the data value of $i$*. There is an immediate correspondence between data strings (that is, sequences of (symbol,value) pairs) and 1-attributed words. Thus, we use in this paper automata and logics that were introduced for data strings also for 1-attributed words.

Attributed $\omega$-words are defined accordingly.

For $i, j \in \mathbb{N}$ with $i \leq j$ we denote the interval $\{i, i+1, \ldots j\}$ by $[i, j]$. As usual we use round brackets to denote open intervals, e.g., $[3, 5) = \{3, 4\}$.

---

[2]As we will use $\mathcal{A}$ for automata we use $\mathcal{V}$ here: **V**ariables.

## 2.2 Basic Data LTL

The logic Basic Data LTL (abbreviated: BD-LTL) has two main types of formulas, *position formulas* and *class formulas*, where, intuitively, class formulas express properties of class words. We first state the syntax of the logic and give an intuitive explanation of its non-standard features afterwards.

We fix a finite set $\mathcal{P} \subseteq \mathcal{PROP}$ of propositions and a finite set $\mathcal{V} \subseteq \mathcal{ATT}$ of attributes.

The syntax of *position formulas* $\varphi$ and *class formulas* $\psi$ of BD-LTL (over $\mathcal{P}$ and $\mathcal{V}$) are defined as follows.

$$\varphi \quad ::= \quad p \mid \neg\varphi \mid \varphi \vee \varphi \mid \mathrm{X}\varphi \mid \mathrm{Y}\varphi \mid \varphi \mathrm{U}\varphi \mid \varphi \mathrm{S}\varphi \mid \mathrm{C}^\delta_{@a}\psi$$
$$\psi \quad ::= \quad \varphi \mid @a \mid \neg\psi \mid \psi \vee \psi \mid \mathrm{X}^=\psi \mid \mathrm{Y}^=\psi \mid \psi \, \mathrm{U}^= \psi \mid \psi \mathrm{S}^=\psi$$

Here, $p \in \mathcal{P}$, $a \in \mathcal{V}$, $\delta \in \mathbb{Z}$. Intuitively, the quantifier $\mathrm{C}_{@a}\psi$ restricts the evaluation of $\psi$ to the class word induced by attribute $a$ at the current position.

Next we define the formal semantics of position formulas. Let $w$ be an attributed word and $i$ a position on $w$:

- $w, i \models p$ if $p \in w[i].\mathcal{P}$;

- $w, i \models \neg\varphi$ if $w, i \not\models \varphi$;

- $w, i \models \varphi_1 \vee \varphi_2$ if $w, i \models \varphi_1$ or $w, i \models \varphi_2$;

- $w, i \models \mathrm{X}\varphi$ if $i + 1 \leq |w|$ and $w, i + 1 \models \varphi$;

- $w, i \models \varphi_1 \mathrm{U}\varphi_2$ if there exists a $j \geq i$ such that $w, j \models \varphi_2$ and $w, j' \models \varphi_1$ for all $j' \in [i, j)$;

- $w, i \models \mathrm{C}^\delta_{@a}\psi$ if $w[i].@a \neq nil$, $i + \delta \in [1, |w|]$, and $w, i + \delta, w[i].@a \models \psi$.

The operators Y and S are the past counterparts of X and U respectively. Their semantics is defined analogously[3].

Next, we define the semantics of class formulas. Let $w$ be an attributed word, $i$ a position on $w$ and $d$ a data value.

- $w, i, d \models \varphi$ if $w, i \models \varphi$;

- $w, i, d \models @a$ if $w[i].@a = d$;

- $w, i, d \models \mathrm{X}^=\varphi$ if there exists a $j \in pos_d(w)$ with $j > i$, and for the smallest such $j$ it holds $w, j, d \models \varphi$;

- $w, i, d \models \varphi_1 \, \mathrm{U}^= \varphi_2$ if there exists a $j \in pos_d(w)$ with $j \geq i$ such that $w, j, d \models \varphi_2$ and $w, k, d \models \varphi_1$ for all $k \in pos_d(w) \cap [i, j)$.

---
[3]To avoid ambiguity: $p\mathrm{S}q$ holds if there is a $q$-position in the past and at the intermediate positions $p$ holds.

For the past class operators Y and S the semantics is defined analogously and the semantics of the Boolean connectors is as usual. Finally, $w \models \varphi$, if $w, 1 \models \varphi$. We denote the set of positional formulas by BD-LTL.

Besides $\bot$ and $\top$ we use the following usual abbreviations:

$$\mathrm{F}\varphi := \top \mathrm{U}\varphi \quad \mathrm{G}\varphi := \neg \mathrm{F}\neg\varphi \quad \mathrm{P}\varphi := \top \mathrm{S}\varphi \quad \mathrm{H}\varphi := \neg \mathrm{P}\neg\varphi$$

The abbreviations $\mathrm{F}^=$ and $\mathrm{G}^=$ and their past counterparts are defined analogously. Furthermore, we abbreviate $\mathrm{C}^\delta_{@a}@b$ by $@a = \mathrm{X}^\delta @b$.

### 2.3 Example: a simple client/server scenario

The following example illustrates how properties can be expressed in BD-LTL.

Consider an internet platform that uses $m$ servers $S_1, \ldots, S_m$ to process queries from clients. Every client shall have a unique client number. As we do not know beforehand how many clients will use the platform, we model the client numbers by the set $\mathcal{D} = \mathbb{N}$.

Each of the servers can either idle, be queried by a client or serve the answer for a query. For server $j$, the actions are modelled by the set of propositions $\{q_j, s_j, i_j\}$. Runs of the internet platform can now be represented by an attributed word with attribute set $\mathcal{ATT} = \{S_1, \ldots, S_m\}$ and set of propositions $\bigcup_{1 \leq j \leq m} \{q_j, s_j, i_j\}$. That a server $S_j$ shall perform exactly one action from $\{q_j, s_j, i_j\}$ at any given time, can be easily expressed by a BD-LTL-formula.

Let us look at an example system with three servers $A$, $B$ and $C$. An example run represented as an attributed word could look as follows.

| Pos | 1 | 2 | 3 | 4 | 5 | 6 |
|---|---|---|---|---|---|---|
| Props | $\{q_A, q_B, i_C\}$ | $\{q_A, q_B, q_C\}$ | $\{s_A, q_B, s_C\}$ | $\{s_A, s_B, i_C\}$ | $\{i_A, s_B, q_C\}$ | $\{i_A, s_B, s_C\}$ |
| A | 1 | 2 | 2 | 1 | – | – |
| B | 2 | 3 | 4 | 2 | 3 | 4 |
| C | – | 1 | 1 | – | 2 | 2 |

Here, e.g., at position 5 server $A$ is idling, server $B$ is serving client 3 and server $C$ is queried by client 2. Properties of runs can be expressed by BD-LTL formulas:

- A client can query a second time on a server only after the first query has been served:

$$\bigwedge_{Z \in \{A,B,C\}} \mathrm{G}(q_Z \to \mathrm{C}_{@Z}((@Z \to \neg q_Z) \, \mathrm{U}^= (@Z \wedge s_Z)))$$

- A server $Z$ can serve a client only if there is an unanswered query by that client (i.e. the last action by that client on $Z$ was a query):

$$\bigwedge_{Z \in \{A,B,C\}} \mathrm{G}(s_Z \to \mathrm{C}_{@Z}((\neg @Z)\mathrm{S}^=(@Z \wedge q_Z))))$$

- A client with an open query on server $A$ shall only be allowed to query server $C$ until server $A$ answered the query:

$$\mathrm{G}(q_A \to \mathrm{C}_{@A}(\neg @B \wedge \mathrm{X}^=((\neg(q_A \wedge A) \wedge \neg(q_B \wedge B)) \, \mathrm{U}^= s_A)))$$

# 3 Expressiveness of BD-LTL

In this section we will give a short overview of established logics on strings with data values and outline how BD-LTL fits in. We give a short introduction to freeze LTL and CLTL$^\diamond$, see [12] and [10] for more details. Afterwards we compare these two logics to BD-LTL.

## 3.1 BD-LTL versus LTL$_1^\downarrow$

*Freeze LTL* is an extension of LTL for data words by a freeze quantifier that binds the data value of the current position to a variable (aka register) and allows to compare the value of a position with the value bound to a variable. Satisfiability for freeze LTL is undecidable even for two registers [12], therefore [12] proposed the 1-register fragment LTL$_1^\downarrow$. In the framework of 1-attributed words, formulas of LTL$_1^\downarrow$ are of the form

$$\varphi ::= p \mid \downarrow\varphi \mid \uparrow \mid \neg\varphi \mid \varphi \wedge \varphi \mid \mathrm{X}\varphi \mid \mathrm{Y}\varphi \mid \varphi\mathrm{U}\varphi \mid \varphi\mathrm{S}\varphi.$$

The formal semantics of LTL$_1^\downarrow$ (on data strings) can be found in [12]. We illustrate it by a simple example: the formula $\mathrm{G}(p \to (\downarrow \mathrm{F}(q \wedge \uparrow)))$ expresses that each $p$-position has a future $q$-position with the same data value.

In [12], the fragment simple LTL$_1^\downarrow$ was invented, where at most one temporal operator is allowed between the the freeze quantifier $\downarrow$ and a value test $\uparrow$. Furthermore, only the unary temporal operators $\mathrm{X}^k, \mathrm{Y}^k, \mathrm{X}^k\mathrm{F}, \mathrm{Y}^k\mathrm{P}$, $k \in \mathbb{N}$ are allowed. Here, $\mathrm{X}^k\mathrm{F}$ is considered a single operator, that is $\downarrow\mathrm{X}^k\mathrm{F}\uparrow$ is an allowed formula. The relative expressive power of BD-LTL and LTL$_1^\downarrow$ can be summarized in the following two propositions.

**Proposition 1** *Every property of 1-attributed words that is expressible in BD-LTL can also be expressed in LTL$_1^\downarrow$.*

The statement also holds for all extensions of BD-LTL considered in Section 5. Note however, that LTL$_1^\downarrow$ is undecidable whereas BD-LTL and its main extension in Section 5 are decidable.

**Proposition 2** *The following logics are equivalent on 1-attributed words*

(i) *Simple LTL$_1^\downarrow$*

(ii) *BD-LTL without Until and Since extended by $\mathrm{F}_{\neq}^\delta$ and $\mathrm{P}_{\neq}^\delta$.*

Here, $\mathrm{F}_{\neq}^\delta\varphi$ intuitively navigates to a future position of distance $\geq \delta$ with a different data value and evaluates $\varphi$ there. In the notation of Section 5 it is an abbreviation for $\top\mathrm{U}_{@a}^\delta(\overline{@a} \wedge \varphi)$. Note, that an analogous operator $F_{=}^\delta\varphi$ for equal data values can be simulated by $\mathrm{C}_{@a}^\delta F^=\varphi$. The proof of both propositions is straightforward and therefore omitted.

## 3.2 BD-LTL versus CLTL$^\diamond$

*Temporal logic of repeating values (CLTL$^\diamond$)* was introduced in [10]. CLTL$^\diamond$-formulas are of the form $\varphi ::= x = X^\delta y \mid x = \diamond y \mid \varphi \wedge \varphi \mid \neg\varphi \mid X\varphi \mid \varphi U\varphi \mid Y\varphi \mid \varphi S\varphi$, where $x, y$ are from a set of variables. A CLTL$^\diamond$-formula with variables $\{x_1, \ldots, x_m\}$ is evaluated on sequences of $m$-tuples of data values (without labels from a finite set) but the extension to $\{x_1, \ldots, x_m\}$-complete attributed strings is straightforward. A formula $x = X^\delta y$ tests whether component $x$ of the current position has the same data value as component $y$ of the $\delta$-next position. A formula $x = \diamond y$ is true if there is a (strict) future position with the same data value on component $y$ as the current position has on component $x$. The semantics of all other operators is as usual.

The following proposition is straightforward, since $x = \diamond y$ and $x = X^\delta y$ can be encoded by $C^0_{@x} X^= F^= @y$ and $C^\delta_{@x} @y$, respectively.

**Proposition 3** *On $\{x_1, \ldots, x_m\}$-complete attributed words BD-LTL is strictly more expressive than CLTL$^\diamond$.*

# 4 Decidability of Basic Data LTL

In this section we show that satisfiability for BD-LTL over attributed words is decidable. The proof is by a reduction to the satisfiability problem for BD-LTL over 1-attributed words (Subsection 4.2) and a reduction of the latter to non-emptiness of data automata (Subsection 4.1). In Subsection 4.3 we extend our decidability result to $\omega$-words. In Subsection 4.4 we show undecidability for two extensions of BD-LTL.

## 4.1 Basic Data LTL over 1-attributed words is decidable

The proof of decidability of satisfiability over 1-attributed words uses register automata (we follow [3]) and data automata [5]. Non-emptiness for data automata is decidable [5]. They are strictly more expressive than register automata and for every register automaton an equivalent data automaton can be effectively constructed [3]; furthermore data automata are closed under intersection, union and letter-to-letter-projection [5].

**Register Automata**

In [17] Kaminski and Francez introduced *Finite-Memory Automata* which work on sequences of data values only but their generalization to 1-attributed words is straightforward. These automata have later been studied in [12] and [22] where they are called *Register automata*. Register automata are equipped with a constant number of registers in which they can store data values which can later be compared with the data value of the current position. We refer to the definition in [3].

A register automaton over a finite alphabet $\Sigma$ is a tuple $\mathcal{R} = (\Sigma, S, s_0, k, \tau_0, \pi, F)$, where Q is a finite set of states, $s_0$ is the initial state, $k$ is the number of registers, $\tau_0 : \{1, ..., k\} \to \{\bot\}$ is the initial register assignment ($\bot \notin \mathcal{D}$ indicates an empty assignment), $\pi$ is a finite set of transitions and $F \subseteq S$ are the accepting states. The transition set $\pi$ consists of *compare transitions* $(i, s, \sigma) \to s'$ and *store transitions* $(s, \sigma) \to (s', i)$, for $i \in \{1, ..., k\}, s, s' \in S$, and $\sigma \in \Sigma$.

A configuration of $\mathcal{R}$ is a pair $(s, \tau)$, where $s \in S$ and $\tau : \{1, ..., k\} \to \mathcal{D} \cup \{\bot\}$ is a register assignment. The initial configuration is $(s_0, \tau_0)$. A *read transition* $(i, s, \sigma) \to s'$ can be applied if the current state is $s$, the next input set is $\sigma$ and the next input data value is already stored in register $i$. It takes the automaton from configuration $(s, \tau)$ to $(s', \tau)$. A write transition $(s, \sigma) \to (s', i)$ can be applied if the current state is $s$, the next input symbol is $\sigma$ and the next input data value $d$ is currently not stored in any register. It takes the automaton from a configuration $(s, \tau)$ to $(s', \tau')$, where $\tau'(i) = d$, and $\tau'(j) = \tau(j)$ for all $j \neq i$.

A run of register automaton over $\Sigma = 2^{\mathcal{P}}$ with $\mathcal{P} \subseteq \mathcal{PROP}$ on an 1-attributed word $w$ with $|w| = n$ is a sequence $(s_0, \tau_0), \ldots, (s_n, \tau_n)$ of configurations, defined in the obvious way. An attributed word $w$ is accepted by a register automaton $\mathcal{R}$ if there exists a run $(s_0, \tau_0), \ldots, (s_n, \tau_n)$ of $\mathcal{R}$ on $w$ with $s_n \in F$.

**Data Automata**

*Data automata* were introduced in [5] and have later been studied in, e.g., [3].

A data automaton $\mathcal{A} = \langle \mathcal{B}, \mathcal{C} \rangle$ over a finite alphabet $\Sigma$ consists of a *base automaton* $\mathcal{B}$ and a *class automaton* $\mathcal{C}$.

- $\mathcal{B} = \langle \Sigma, \Gamma_\mathcal{B}, S_\mathcal{B}, s_{0\mathcal{B}}, \pi_\mathcal{B}, F_\mathcal{B} \rangle$ is a nondeterministic letter-to-letter string transducer with input alphabet $\Sigma$, output alphabet $\Gamma_\mathcal{B}$, an initial state $s_{0\mathcal{B}} \in S_\mathcal{B}$, a transition relation $\pi_\mathcal{B} \subseteq S_\mathcal{B} \times \Sigma \times \Gamma_\mathcal{B} \times S_\mathcal{B}$ and a set $F_\mathcal{B} \subseteq S_\mathcal{B}$ of accepting states. In each step $\mathcal{B}$ replaces the current label $\sigma \in \Sigma$ with a symbol $\gamma \in \Gamma_\mathcal{B}$.

- $\mathcal{C} = \langle \Gamma_\mathcal{B}, Q_\mathcal{C}, s_{0\mathcal{C}}, \pi_\mathcal{C}, F_\mathcal{C} \rangle$ is a nondeterministic string automaton with input alphabet $\Gamma_\mathcal{B}$, an initial state $s_{0\mathcal{C}} \in Q_\mathcal{C}$, a transition relation $\pi_\mathcal{C} \subseteq S_\mathcal{C} \times \Gamma_\mathcal{B} \times S_\mathcal{C}$ and a set $F_\mathcal{C} \subseteq S_\mathcal{C}$ of accepting states.

An 1-attributed word $w$ of length $n$ is accepted by $\mathcal{A} = \langle \mathcal{B}, \mathcal{C} \rangle$ over $\Sigma = 2^{\mathcal{P}}$ where $\mathcal{P} \subseteq \mathcal{PROP}$ if there is an accepting run of $\mathcal{B}$ on $str(w)$, yielding an output string $\gamma_1 \ldots \gamma_n$, such that, for each set $pos_w(d) = \{i_1, ..., i_k\} \subseteq \{1, ..., n\}$ of class positions with $d$ occurring in $w$ and $i_1 < \ldots < i_k$ the class automaton accepts $\gamma_{i_1} \ldots \gamma_{i_k}$.

We denote the set of words accepted by a data or register automaton $\mathcal{A}$ by $\mathcal{L}(\mathcal{A})$.

**Theorem 1** *Satisfiability for BD-LTL on 1-attributed words is decidable.*

**Proof.** Let $\varphi$ be a BD-LTL formula over a proposition set $\mathcal{P}$ and the attribute set $\{a\}$.

In the following we often call 1-attributed words simply *words*. Our automata will expect instead of words $w$ over P extended words $w'$ with additional propositions. First, $w'$ allows the subformulas of $\varphi$ as propositions. The intention is that a position $i$ of $w'$ is marked with $\psi$ if and only if $w, i \models \psi$. Furthermore, we use propositions $=_r$ for every $r \in \{-N, \ldots, -1, 1, \ldots, N\}$, for some $N$ that is at least as large as every $\delta$ occurring in $\varphi$. Proposition $=_r$ shall hold at position $i$ if and only if $w[i].@a = w[i+r].@a$. Whether these propositions are correct in a given $w'$ can be easily tested by a register automaton.

We call an extended word $w'$ *valid* if the propositions $=_r$ are as intended and each position $i$ carries exactly those formulas holding at $i$ in $w$. The word $w'$ is *valid with respect to a subformula $\psi$* of $\varphi$ if $\psi$ is a proposition or $w'$ is consistent with respect to the maximal strict subformulas of $\psi$. As an example, a word is valid with respect to $\varphi = \varphi_1 U \varphi_2$ if for each $\varphi$-position $i$ there is a $\varphi_2$-position $j \geq i$ such that all positions $k$, $i \leq k < j$ are marked by $\varphi_1$.

Clearly, each word $w$ has a unique valid extension. Thus, $\varphi$ is satisfiable if and only if it has a valid extension $w'$ in which position 1 carries the proposition $\varphi$.

We show in the following that from a BD-LTL formula a data automaton $\mathcal{A}_\varphi$ can be constructed that checks whether an extended word $w'$ is valid. The statement of the theorem then follows from the decidability of non-emptiness of data automata.

To this end, we construct for each subformula $\psi$ of $\varphi$ a data (or register) automaton $\mathcal{A}'_\psi$ that decides whether a given word is valid with respect to $\psi$ under the assumption that it is valid with respect to all strict subformulas of $\psi$. By intersecting all these automata and intersecting the result with a register automaton that tests that the $=_r$ propositions are correct and an automaton that tests whether position 1 carries $\varphi$, we obtain $\mathcal{A}_\varphi$.

The construction of $\mathcal{A}'_\psi$ is straightforward for formulas of the types $p \mid \neg \varphi \mid \varphi \vee \varphi \mid X\varphi \mid Y\varphi \mid \varphi U \varphi \mid \varphi S \varphi$. Basically, these automata do not need a class automaton. The construction is equally straightforward for all types of class formulas. In these cases, basically only class automata are needed. To deal with the $\delta$-shift in formulas of the form $C^\delta_{@a}\psi$ we use the propositions $=_r$. E.g., to validate with respect to $\psi = C^7_{@a} F^= \chi$ at position $i$, the class automaton of $\mathcal{A}'_\psi$ infers from the $=_r$ propositions how many positions the class word has between $i$ and $i + 7$, then it skips these positions and starts searching for a $\chi$-position from there. □

## 4.2 Basic Data LTL is decidable

**Theorem 2** *Satisfiability for BD-LTL is decidable.*

**Proof.** Thanks to Theorem 1 it suffices to reduce the satisfiability problem for BD-LTL to the satisfiability problem for BD-LTL on 1-attributed words. That is, for a given BD-LTL formula $\chi$, we construct a BD-LTL formula $\chi'$ such that $\chi'$ holds for some 1-attributed word if and only if $\chi$ holds for some attributed word. Therefore we use an encoding of attributed words by 1-attributed words that has been introduced in [13].

|       | Pos    | 1     | 2     | 3     |
|-------|--------|-------|-------|-------|
| $w =$ | Props  | $p$   | $q$   | $p,q$ |
|       | $\text{att}_1$ | $d_1$ | –     | $d_3$ |
|       | $\text{att}_2$ | –     | $d_2$ | $d_4$ |

|        | Pos   | 1     | 2     | 3     | 4     | 5     | 6     |
|--------|-------|-------|-------|-------|-------|-------|-------|
| $w' =$ | Props | $p$ $\text{att}_1$ $R$ | $p$ $\text{att}_2$ | $q$ $\text{att}_1$ | $q$ $\text{att}_2$ $R$ | $p,q$ $\text{att}_1$ $R$ | $p,q$ $\text{att}_2$ $R$ |
|        | $a$   | $d_1$ | $d_2$ | $d_3$ | $d_2$ | $d_3$ | $d_4$ |

Figure 1: How an attributed word $w$ with two attributes is encoded as 1-attributed word $w'$.

Let $\mathcal{P}$ and $\mathcal{V} = \{a_1, \ldots, a_m\}$ be the set of propositions and the set of attributes used in $\chi$, respectively. The set of propositions used in $\chi'$ is $\mathcal{P} \cup \mathcal{V} \cup \{R\}$.

The intuition is that every position $i$ of a model $w$ of $\chi$ is encoded by a block of $m$ positions in a model $w'$ of $\chi'$. Position $j$ in the $i$-th block of $w'$ thus represents attribute $a_j$ of the $i$th position of $w$. However, as some positions in $w$ might not carry values for all attributes, the positions corresponding to existing attributes are marked by proposition $R$ (see Figure 1 for an example).

The formula $\chi'$ constructed from $\chi$ is of the form $\varphi_{\text{structure}} \wedge t(\chi)$ where

- $\varphi_{\text{structure}}$ makes sure that the 1-attributed word is well-formed, that is, it is a concatenation of blocks of length $m$, where, in each block, the $j$-th position has proposition $\text{att}_j$ (and no other $\text{att}_k$) and all positions carry the same propositions from $\mathcal{P}$, and

- $t(\chi)$ is obtained from $\chi$ by a straightforward inductive construction.

The translation $t$ makes use of additional translations $t_i$ and $t_{\max}$. More precisely,
$$t_i(\varphi) = \bigvee_{j=1}^{m} (\text{att}_j \to X^{i-j}\varphi),$$
and
$$t_{\max}(\varphi) = \bigvee_{i=1}^{m} \left( \text{att}_i \to \bigvee_{\delta=0}^{m-i} \left( (@a = X^\delta @a \wedge \bigwedge_{\delta'=\delta+1}^{m-i} \neg @a = X^{\delta'} @a) \to X^\delta \varphi \right) \right),$$

Here, we use $@a = X^\delta @a$ as an abbreviation for $C^\delta_{@a}(R \wedge @a)$. Intuitively, $t_i(\varphi)$ is a formula which first navigates to the $i$-th position of the current block and evaluates $\varphi$. Likewise, $t_{\max}(\varphi)$ is a formula which first navigates to the last position of the current block with the same (active) data value and evaluates $\varphi$.

The inductive translation $t$ of positional formulas $\chi$ is defined by

- $t(p) := p$ for all $p \in \mathcal{P}$

- $t(\neg\varphi) := \neg t(\varphi)$
- $t(\varphi \vee \psi) := t(\varphi) \vee t(\psi)$
- $t(\mathrm{X}\varphi) := \mathrm{X}^m t(\varphi)$
- $t(\varphi \mathrm{U} \psi) := t(\varphi)\mathrm{U}t(\psi)$
- $t(@a_i = \mathrm{X}^k @a_j) := t_i(@a = X^{km+(j-i)}@a)$
- $t(\mathrm{C}_{@a_i}\varphi) := t_i(R \wedge t(\varphi))$

for all $i, j \in \{1, \ldots, m\}$. Translations of the positional formulas for the past are analogous.

Now we give the translation for class formulas. Here, the basic idea is that each navigation step ends at a position with the currently frozen data value.

- $t(@a_j) := \bigvee_{i=1}^m (\mathrm{att}_i \to (@a = X^{j-i}@a))$
- $t(\neg\varphi) = \neg t(\varphi)$
- $t(\varphi \vee \psi) = t(\varphi) \vee t(\psi)$
- $t(\mathrm{X}^=\varphi) = t_{\max}(\mathrm{X}^=(\neg R\, \mathrm{U}^=(R \wedge t(\varphi))))$
- $t(\varphi\, \mathrm{U}^= \psi) = (R \to t(\varphi))\, \mathrm{U}^=(R \wedge t(\psi))$

The translations for the past operators are analogous. The correctness of the reduction is straightforward.

☐

The complexity of the satisfiability problem for BD-LTL over 1-attributed words is probably very bad. As for two-variable logics [5], it is (efficiently) interreducible with the reachability problem for Petri Nets [15].

### 4.3 BD-LTL on infinite attributed words is decidable

Over attributed $\omega$-words BD-LTL remains decidable. First, we observe that the mapping used in Theorem 2 works also for the infinite case, i.e. the satisfiability problem for BD-LTL over attributed $\omega$- words is reducible to the satisfiability problem over 1-attributed $\omega$-words. Hence, it remains to prove the decidability for the latter problem. We do this by a reduction to the non-emptiness problem for data $\omega$-automata which is shown to be decidable in [5]. These automata are defined analogously to data automata. We only describe the differences here. A data $\omega$-automaton $\mathcal{A} = \langle \mathcal{B}_{inf}, \mathcal{C}, \mathcal{C}_{inf} \rangle$ consists of a base automaton $\mathcal{B}_{inf}$ which is a Büchi letter-to-letter transducer with output over some alphabet $\Gamma$, a finitary class automaton $\mathcal{C}$ which is a finite string automaton over $\Gamma$ and an infinitary class automaton $\mathcal{C}_{inf}$, which is another Büchi automaton over $\Gamma$. An 1-attributed $\omega$-word $w$ is accepted if the base automaton has an accepting run over the string projection of $w$ with output $\gamma_1\gamma_2\ldots$ such that for every finite class $i_1 < \ldots < i_m$ the string $\gamma_{i_1}\ldots\gamma_{i_m}$ is accepted by $\mathcal{B}$ and for every infinite class $i_1 < i_2 < \ldots$, the $\omega$-string $\gamma_{i_1}\gamma_{i_2}\ldots$ is accepted by $\mathcal{B}_{inf}$.

**Theorem 3** *Satisfiability for BD-LTL on attributed $\omega$-words is decidable.*

**Proof.** (Sketch) The proof is by reduction to the non-emptines problem for data $\omega$-automata and is along very similar lines as that of Theorems 1 and 2. By applying the mapping given in Theorem 2 we first make a reduction to the satisfiability problem for BD-LTL on 1-attributed $\omega$-words. Then, given a BD-LTL formula $\varphi$ defined on $\{a\}$-complete attributed words we can construct a data $\omega$-automaton which is non-empty if and only if $\varphi$ is satisfiable. The automaton $\mathcal{A}'_\psi$ that checks whether a given word is valid with respect to a formula $\psi$ differs from its counterpart in the finite case, basically, only for $\psi = \varphi_1 \mathrm{U} \varphi_2$ or $\psi = \varphi_1 \mathrm{U}^= \varphi_2$. The difficulty is to ensure that whenever $\psi$ occurs the formula $\varphi_2$ definitely occurs in future. But these cases can be handled easily using classical techniques in automata construction for temporal logics on $\omega$-words [25]. For a formula $\psi = \varphi_1 \mathrm{U} \varphi_2$ the automaton $\mathcal{A}'_\psi$ checks that the input word has a suffix where either positions labeled with $\psi_1 \mathrm{U} \varphi_2$ and $\varphi_2$ or positions labeled with $\neg(\varphi_1 \mathrm{U} \varphi_2)$ occur infinitely often. The acceptance condition for the class operator $\mathrm{U}^=$ is defined analogously. □

## 4.4 Undecidable Extensions

Extensions of BD-LTL quickly yield undecidability. We consider two such extensions here.

### BD-LTL with Navigation along Tuples

We extend $\mathrm{C}_{@a}$ to a quantifier $\mathrm{C}_{@a,@b}$ that 'freezes' the values $d_a$ and $d_b$ of the attributes $a$ and $b$, respectively. Operators $\mathrm{X}^=, \mathrm{Y}^=, \mathrm{U}^=$ and $\mathrm{S}^=$ in the scope of $\mathrm{C}_{@a,@b}$ then move along positions that have attributes with data values $d_a$ and $d_b$. At such positions the values of tuples of attributes can be tested for equality with $(d_a, d_b)$. For example the property 'there is a future position with proposition $p$ where attribute $c$ carries the same data value as attribute $a$ at the current position, likewise for $d$ and $b$' can be expressed by $\mathrm{C}_{@a,@b} F^=((@c, @d) \wedge p)$.

However, already a restricted version of this extension is undecidable. We consider the operators $\mathrm{X}_{@a,@b}$ and $\mathrm{Y}_{@a,@b}$. Let the semantics of $\mathrm{X}_{@a,@b}$ be defined by $w, i \models \mathrm{X}_{@a,@b} \varphi$ if there is a $j > i$ with $w[i].@a = w[j].@a$ and $w[i].@b = w[j].@b$ and for the smallest such $j$ it holds $w, j \models \varphi$. The operator $\mathrm{Y}_{@a,@b}$ is defined analogously.

**Theorem 4** *BD-LTL extended by the operators $\mathrm{X}_{@a,@b}$ and $\mathrm{Y}_{@a,@b}$ is undecidable on finite (or infinite) attributed words.*

**Proof.** The proof is along the lines of Proposition 27 in [5]. For the convenience of the reader, we sketch the proof.

We reduce from the Post Correspondence Problem (PCP) which is defined by

*Problem:* PCP
*Input:* $(u_1, v_1), \ldots, (u_k, v_k) \in \Sigma^* \times \Sigma^*$
*Question:* Is there a non-empty sequence $(u_{i_1}, v_{i_1}), \ldots, (u_{i_n}, v_{i_n})$ such that $u_{i_1} \ldots u_{i_n} = v_{i_1} \ldots v_{i_n}$?

Given a PCP $P$ we construct a BD-LTL formula $\varphi_P$ using the operators $X_{@a,@b}$ and $Y_{@a,@b}$ such that there is a valid sequence for $P$ if and only if $\varphi_P$ is satisfiable. We can assume w.l.o.g. that if there is a valid sequence for $P$, then there is one of odd length.

Let $P := (u_1, v_1), \ldots, (u_k, v_k) \in \Sigma^* \times \Sigma^*$. Then $\varphi_P$ uses the propositions $\Sigma \times \bar{\Sigma}$ where $\bar{\Sigma} := \{\bar{\sigma} \mid \sigma \in \Sigma\}$.

A valid sequence $(u_{i_1}, v_{i_1}) \ldots (u_{i_n}, v_{i_n})$ of odd length for $P$ is mapped to an attributed word $w$ such that:

(1) Every position of $w$ bears exactly one proposition and the word projection of $w$ is of the form $u_{i_1} \bar{v}_{i_1} \ldots u_{i_n} \bar{v}_{i_n}$.

(2) The attributes $a$ and $b$ are present at all positions and the data values of the attributes $a$ and $b$ for the word $u := u_{i_1} \ldots u_{i_n}$ are of the form $(d_1^a, d_1^b)(d_1^a, d_2^b)(d_2^a, d_2^b), \ldots, (d_{n-1}^a, d_n^b), (d_n^a, d_n^b)$. The same for the word $\bar{v} := \bar{v}_{i_1} \ldots \bar{v}_{i_n}$.

(3) In $u := u_{i_1} \ldots u_{i_n}$ all values of attribute $a$ occur exactly twice, except for the value of $a$ at the last position of $u$. Similarly all values of attribute $b$ occur exactly twice, except for the value of $b$ at the first position of $u$. The same holds for $\bar{v} := \bar{v}_{i_1} \ldots \bar{v}_{i_n}$.

(4) Every pair $(d_a, d_b)$ of data values for the attributes $a$ and $b$ occurs exactly twice, once in $u$ and once in $\bar{v}$ and the position in $u$ is labeled by $\sigma$ if and only if the corresponding position in $\bar{v}$ is labeled by $\bar{\sigma}$.

Note that conditions (2)-(4) ensure that $u = v'$, where $v'$ results from $\bar{v}$ by replacing every $\bar{\sigma}$ in $\bar{v}$ by $\sigma$.

The first three properties can be checked easily by a BD-LTL formula. Condition (4) can be checked using the additional operators:

$$G\bigg(\bigvee_{\sigma \in \Sigma} \Big(\sigma \to \big((X_{@a,@b}\bar{\sigma} \wedge \neg X_{@a,@b} X_{@a,@b}\top \wedge \neg Y_{@a,@b}\top) \vee$$
$$(Y_{@a,@b}\bar{\sigma} \wedge \neg Y_{@a,@b} Y_{@a,@b}\top \wedge \neg X_{@a,@b}\top)\big)\Big)\bigg)$$

$$\wedge$$

$$G\bigg(\bigvee_{\bar{\sigma} \in \bar{\Sigma}} \Big(\bar{\sigma} \to \big((X_{@a,@b}\sigma \wedge \neg X_{@a,@b} X_{@a,@b}\top \wedge \neg Y_{@a,@b}\top) \vee$$
$$(Y_{@a,@b}\sigma \wedge \neg Y_{@a,@b} Y_{@a,@b}\top \wedge \neg X_{@a,@b}\top)\big)\Big)\bigg) \quad \square$$

**BD-LTL with From-Now-On Operator**

The *from-now-on*-operator N introduced in [18] restricts the range of past operators. For an attributed word $w = w_1 \ldots w_n$ and a position $i$ of $w$ let $suf_i(w) := w_i \ldots w_n$ be the suffix of $w$ starting at position $i$. The semantics of N is then defined by

- $w, i \models N\varphi$ if $suf_i(w), 1 \models \varphi$

**Theorem 5** *BD-LTL extended by the operator* N *is undecidable on finite (or infinite) attributed words.*

**Proof.** For technical reasons, we prove that BD-LTL with the operator $\bar{N}$, the dual on N is undecidable. For an attributed word $w = w_1 \ldots w_n$ and a position $i$ of $w$ let $pre_i(w) := w_1 \ldots w_i$ be the prefix of $w$ ending at position $i$. The semantics of $\bar{N}$ is then defined by

- $w, i \models \bar{N}\varphi$ if $pre_i(w), i \models \varphi$

As BD-LTL is symmetric with respect to the availability of future and past operators, undecidability of BD-LTL with N follows from the undecidability of BD-LTL with $\bar{N}$.

The proof is by a reduction from the nonemptiness problem for Minsky two counter automata [21].

A *Minsky 2-CA* is a finite automaton equipped with two counters $C_1$ and $C_2$. With each transition of the automaton a *counter action* is associated. In a counter action a counter can be incremented or decremented by 1 ($inc_1$, $inc_2$, $dec_1$, $dec_2$) or counter can be tested for 0 ($ifzero_1$, $ifzero_2$). Initially, both counters have the value 0. The semantics is straightforward. If a counter $i$ has value 0 a transition with a $dec_i$-action can not be applied. A transition with a $ifzero_i$ action can only be applied if counter $i$ is 0.

It is well known that the emptiness problem for Minsky 2-CA is undecidable [21].

Our reduction constructs, for every Minsky 2-CA $\mathcal{C}$ an extended BD-LTL formula $\varphi$ such that $L(\mathcal{C}) \neq \emptyset$ if and only if $\varphi$ is satisfiable by a 1-attributed word. We can assume without loss of generality that $\mathcal{C}$ only accepts if both counters are zero.

We encode runs of $\mathcal{C}$ by 1-attributed words as follows. Each position $i$ carries two propositions, one is the state of $\mathcal{C}$ after step $i$ and the other the action in step $i$. As in an accepting run there are, for each $i$ exactly as many $inc_i$ actions as there are $dec_i$ actions we can assign data values in a way such that each data value

- either occurs exactly once at an ifzero-test or
- once at an $inc_i$ action and once at the corresponding[4] $dec_i$-action.

---

[4]That is, if the first increments counter 1 from $m$ to $m+1$ the other is the action that decreases it back from $m+1$ to $m$.

It is not hard to write an BD-LTL formula with the $\bar{\text{N}}$ operator that expresses that

(1) successive positions are consistent with respect to the transitions of $\mathcal{C}$, the first position carries starting state and the final position carries an accepting state,

(2) every data value only occurs once at an ifzero-test or twice: once at a $\text{dec}_i$ action and once at a smaller position with a $\text{inc}_i$-action, and

(3) an $\text{ifzero}_i$-action never occurs between an $\text{inc}_i$-action and the $\text{dec}_i$-action with the same data value.

Note that condition (3) makes sure that an $\text{ifzero}_i$-action only occurs if counter $C_i$ is 0. Conditions (1) and (2) can be easily expressed by BD-LTL.

Condition (3) can be expressed by the formula

$$\bigwedge_{i=1}^{2} \text{G}(\text{ifzero}_i \to \bar{\text{N}}\text{H}(\text{inc}_i \to \text{C}_{@a}\text{F}^=\text{dec}_i))$$

□

## 5 Extended Navigation

As already discussed before, the navigational abilities of BD-LTL are limited. It seemingly cannot[5] even express the simple property that for every $p$-position $i$ there is a $q$-position $j > i$ such that $w[j].@b \neq w[i].@a$. Furthermore, in class formulas $\rho \text{U}^=\tau$, the formula $\rho$ can only refer to positions of the current class. Of course, it would be desirable to allow more general forms of "Until navigation".

In this section we discuss different possibilities to extend the navigational abilities of BD-LTL in an "Until fashion", some of which are decidable and some undecidable. In particular, we exhibit an U-operator with the ability to navigate to a position with a different attribute value and to state some properties on (all) intermediate positions and show that BD-LTL remains decidable with this extension. The property stated in the previous paragraph can be expressed using this operator.

The extensions we study allow formulas of the type $\rho \text{U}^\delta_{@a} \tau$, where $\delta \geq 0$. Intuitively, this operator "freezes" the current value of attribute $@a$ and searches for a position $j$ such that $\tau$ holds at $j$ and $\rho$ hold everywhere in $[i+\delta, j)$. In formulas as above, we will refer to $\rho$ as the *intermediate formula* and $\tau$ as the *target formula*. The "shift" parameter $\delta$ is needed as we aim to design a semantic extension of simple $\text{LTL}_1^\downarrow$.

Syntactically, the formulas $\rho$ and $\tau$ are positive Boolean combinations of position formulas and positive and negative attribute tests. More formally, we define the syntax of *U-subformulas* $\chi$ by $\chi ::= \varphi \mid @b \mid \overline{@b} \mid \chi \vee \chi \mid \chi \wedge \chi$.
Intuitively, negative attribute tests $\overline{@b}$ check that attribute $b$ has a value (!)
that is different from the current frozen value.

---

[5]We did not attempt to find a proof for this statement as we were aiming for an extended logic, anyway. However, we did not find a simple way to express the property.

Thus, the semantics of formulas $\rho U_{@a}^\delta \tau$, where $\rho$ and $\tau$ are *U*-subformulas, is defined by the following additional rules.

- $w, i \models \rho U_{@a}^\delta \tau$ if there exists a $j \geq i + \delta$ such that $w, j, w[i].@a \models \tau$ and $w, k, w[i].@a \models \rho$ for all $k \in [i+\delta, j)$

- $w, i, d \models \overline{@b}$ if $w[i].@b \notin \{nil, d\}$.

We simply use $U_{@a}$ instead of $U_{@a}^0$. We remark that $\rho U_{@a}^{-\delta} \tau$, for $\delta \geq 0$ can be expressed by $(\rho\, U_{@a}\, \tau \wedge \bigwedge_{i=1}^{\delta} \rho_i) \vee (\bigvee_{j=1}^{\delta}(\tau_j \wedge \bigwedge_{i=j+1}^{\delta} \rho_i))$, where, for $k \in [1, \delta]$, $\rho_k$ and $\tau_k$ are obtained from $\rho$ and $\tau$, respectively, by replacing every position formula $\varphi$ by $Y^k \varphi$, every $@b$ by $@a = Y^k @b$ and every $\overline{@b}$ by $\neg @a = Y^k @b$. It can be observed that this formula has the intended meaning (that is, the semantics obtained by using $-\delta$ in the above semantics definition). $\rho S_{@a} \tau$ is defined analogously.

First of all, we will see that the above mentioned restriction for class formulas $\rho U^= \tau$ is indeed crucial. More precisely, if we allow positive attribute tests in the target formula of a formula $\rho\, U_{@a}\, \tau$ then the logic becomes undecidable even if the intermediate formulas are restricted to position formulas.

## 5.1 Extended equality-navigation is undecidable

**Theorem 6** *Let $\mathcal{L}$ denote the extension of BD-LTL by the formation rule $\varphi ::= \chi\, U_{@a}\, \chi$, where $\chi$ denotes U-subformulas such that*

- *all intermediate formulas are position formulas and*
- *all target formulas are of the form $@a \wedge \varphi$ with a position formula $\varphi$.*

*Then, satisfiability of $\mathcal{L}$ on finite (or infinite) attributed words is undecidable. This holds even for 1-attributed words.*

**Proof.** As in Theorem 5, we reduce from the non-emptiness problem for Minsky two counter automata.

As before, conditions (1) and (2) from the proof of Theorem 5 can be easily expressed in BD-LTL. Condition (3) can be expressed by the formula

$$\bigwedge_{i=1}^{2} G(\text{inc}_i \rightarrow (\neg \text{ifzero}_i\, U_{@a}(@a \wedge \text{dec}_i))).$$

A slightly modified reduction works for infinite 1-attributed words. □

## 5.2 Extended inequality navigation

As Theorem 6 does not leave much room for extensions of $U_{@a}$ operators with positive attribute tests in the target formula we focus on negative attribute tests in target formulas. However, as $\rho U_{@a}^\delta(\tau_1 \vee \tau_2) \equiv (\rho U_{@a}^\delta \tau_1) \vee (\rho\, U_{@a}\, \tau_2)$ and position formulas are closed under conjunctions it is clearly sufficient to consider target formulas of the form $\varphi \wedge \overline{@b_1} \wedge \cdots \wedge \overline{@b_k}$. Unfortunately, at this point our techniques can only deal with the case $k = 1$.

We turn our attention now to the intermediate formulas $\rho$. We recall that in the case of positive attribute tests in target formulas even position formulas as intermediate formulas yield undecidability. In the case of (single) negative attribute tests in target formulas we can allow arbitrary intermediate position formulas.

Furthermore, we can add positive and negative attribute tests, but only in a limited way. More precisely, we define the logic XD-LTL by adding $\varphi ::= \chi U_{@a}^\delta \chi' \mid \chi S_{@a}^\delta \chi'$, to the formation rules of BD-LTL and requiring that

1. $\chi$ is restricted to formulas of the form $\rho \vee (@b \wedge \rho^=) \vee (\overline{@b} \wedge \rho^{\neq})$ where $\rho^=, \rho^{\neq}$ are position formulas and $\rho^{\neq}$ logically implies[6] $\rho^=$, and

2. $\chi'$ is restricted to formulas of the form $\overline{@b} \wedge \tau$, where $\tau$ is a position formula.

Intuitively, $\rho^=$ constrains positions where $@b$ equals the current value of $@a$ whereas $\rho^{\neq}$ constrains those where it does not. The requirement that $\rho^{\neq}$ implies $\rho^=$ is needed for the proof of Theorem 8.

Clearly XD-LTL strictly extends BD-LTL and is contained in $\mathrm{LTL}_1^{\downarrow}$. Further it strictly extends two-variable logic on 1-attributed words.

Following the general idea of the decidability proof for BD-LTL we first show decidability of satisfiability for 1-attributed words and reduce the general case to this one.

## 5.3 Satisfiability for XD-LTL

**Theorem 7** *Satisfiability for XD-LTL on finite 1-attributed words is decidable.*

**Proof.** The proof basically extends the proof of Theorem 1 for the new formulas. As usual, we only describe the case of $U_{@a}^\delta$-formulas.

Note that in the case of 1-attributed words, any additional disjunct $\rho$ in the intermediate formula can be pushed into the disjunction by or-ing it with both $\rho^=$ and $\rho^{\neq}$. It thus only remains to show how to construct $\mathcal{A}_\psi'$ for formulas $\psi = (@a \wedge \rho^=) \vee (\overline{@a} \wedge \rho^{\neq}) U_{@a}^\delta (\overline{@a} \wedge \tau)$, where $\rho^=$ and $\rho^{\neq}$ are position formulas and $\rho^{\neq}$ implies $\rho^=$.

Before we describe the construction of $\mathcal{A}_\psi'$, we first analyze the relationships between positions at which such a formula holds. Clearly, $w, i \models (@a \wedge \rho^=) \vee (\overline{@a} \wedge \rho^{\neq}) U_{@a}^\delta (\overline{@a} \wedge \tau)$ if and only if there is a position $j \geq i + \delta$ such that

(I) $w, j \models \tau$,

---

[6] Readers who prefer a syntactical criterion might think of a formula $\rho^=$ of the form $\varphi \vee \rho^{\neq}$.

| Pos | 1 | 2 | 3 | 4 | 5 | 6 | 7 | 8 | 9 | 10 |
|---|---|---|---|---|---|---|---|---|---|---|
| Props | | | $\psi$ | $\psi$ $\tau$ | | $\psi$ | $\psi$ | | | $\tau$ |
| | $\rho^=$ | | $\rho^=$ | $\rho^=$ | $\rho^{\neq}$ | $\rho^=$ | $\rho^{\neq}$ | $\rho^{\neq}$ | $\rho^{\neq}$ | |
| $a$ | 1 | 2 | 1 | 1 | 2 | 1 | 2 | 2 | 2 | 3 |

Figure 2: For $\psi = (@a \wedge \rho^=) \vee (\overline{@a} \wedge \rho^{\neq}) \mathrm{U}^2_{@a} (\overline{@a} \wedge \tau)$, the herd of the shepherd at position 10 is $\{3, 4, 6, 7\}$. Position 4 is no shepherd. Position 6 is a $\rho$-stair for positions $3, 4$. Positions $3, 4$ are $\rho$-far from 10. Positions $3, 4, 6$ are special for position 10, whereas there is no special position for position 4. Further $e^-(10) = 3$ and $e^+(10) = 6$.

(II) $w[j].@a \neq w[i].@a$, and

(III) for the minimal such $j$ every position $k \in [i + \delta, j)$ fulfills

   (a) $w, k \models \rho^{\neq}$ or
   
   (b) $w[k].@a = w[i].@a$ and $w, k \models \rho^=$.

For a given position $i$ with $w, i \models \psi$ we call the position $j$ of criterion (III) *the $\psi$-shepherd for $i$*. We write $H(j)$ for the herd of $j$, that is the set of positions for which $j$ is a $\psi$-shepherd (see Figure 2). A position $j$ with $H(j) \neq \emptyset$ is also called a *shepherd*.

With each $\tau$-position $j$ we associate a set of special positions. They are used to deal with $\psi$-positions that need (IIIb) at least once ($\rho$-special positions) and with cases where the smallest $\tau$-position larger than some $i$ has the same data value as $i$ ($\tau$-special positions).

- If for some $i \in H(j)$ there is a $k \in [i + \delta, j)$ such that $w, k \not\models \rho^{\neq}$ (and thus (IIIb) holds), then we say that $i$ is *$\rho$-far* for $j$ and $k$ is a *$\rho$-stair* for $j$ and both are *$\rho$-special for $j$*.

- If for some $i \in H(j)$ there is a $k \in [i + \delta, j)$ such that $w, k \models \tau$ (and due to minimality of $j$ thus $w[k].@a = w[i].@a$ holds), then we say that $i$ is *$\tau$-far* for $j$ and $k$ is a *$\tau$-stair* for $j$ and both are *$\tau$-special for $j$*.

We write $S(j)$ for the set of ($\rho$- or $\tau$-) special positions for $j$ and remark that $S(j)$ does not need to be a subset of $H(j)$. See Figure 2 for an illustration of these definitions.

For technical reasons, we define the set $S(j)$ also for $\tau$-positions with an empty herd. To this end, for such $\tau$-positions $j$ we put the maximal $k \in [j - \delta, j)$ with $w'[k].@a \neq w'[j].@a$ such that

- all positions in $(k, j)$ are $\rho^{\neq}$-positions, and

- $k$ is a $\tau$-position or it is marked by $\rho^=$ but not by $\rho^{\neq}$

into $S(j)$, if such a position exists. Then we add all positions $\ell \in (j - \delta, k)$ with $w'[\ell].@a = w'[k].@a$ to $S(j)$.

For positions $j$ with non-empty $S(j)$, we denote the minimal and maximal special positions for $j$ by $e^-(j)$ and $e^+(j)$, respectively (see Figure 2) and call $[e^-(j), e^+(j)]$ the *special interval* $I(j)$ of $j$. Note that $e^+(j) < j$ and there is no further $\tau$-position in $(e^+(j) + \delta, j)$ (as such a position would contradict the minimality of $j$ or would be special for $j$), thus $j$ can be easily identified, given $e^+(j)$. If $S(j)$ is empty, then $I(j)$ is the empty interval. By construction, if $H(j) = \emptyset$, the smallest element of $S(j)$ is larger than $j - \delta$, thus $|I(j)| < \delta$.

**Claim 1**

(1a) All positions in $S(j)$ have the same data value.

(1b) For each $\tau$-position $j$ with $S(j) \neq \emptyset$, $e^+(j)$ is the largest position $k < j$ with $w'[k].@a \neq w'[j].@a$ such that

- all positions in $(k, j)$ are $\rho^{\neq}$-positions, and
- $k$ is a $\tau$-position or it is marked by $\rho^=$ but not by $\rho^{\neq}$.

(1c) For each $\tau$-position $j$ with $S(j) \neq \emptyset$, $e^-(j)$ is the smallest position $m \leq e^+(j)$ with $w'[m].@a = w'[e^+(j)].@a$ such that

- all $l \in [m + \delta, j) - S(j)$ are $\rho^{\neq}$-labeled,
- all $l \in [m + \delta, j)$ are $\rho^=$-labeled,
- there is no $\tau$-labeled position $l \in [m + \delta, j) - S(j)$.

(1d) $|I(j) \cap I(j')| \leq \delta$ for $j \neq j'$.

To show Claim (1a), let $k$ be the minimal $\rho$-far position in $S(j)$, in case $H(j) \neq \emptyset$. Clearly, all $\rho$-stairs have the same value as $k$ and in turn all other $\rho$-far positions as well. If $H(j) = \emptyset$ then all positions have the same value by definition of $S(j)$. A similar argument shows that all $\tau$-special positions have the same data value. Let now $k$ be the maximum $\rho$-stair and $l$ the maximum $\tau$-stair in $S(j)$ and let us assume that $w[k].@a \neq w[l].@a$. If $k < l$, $l$ would be a shepherd for the smallest $\rho$-far position in $S(j)$, contradicting the minimality of $j$. If $l < k$, then $w[k].@a \neq w[i].@a$ and $w, k \not\models \rho^{\neq}$ for the smallest $\tau$-far position, again a contradiction.

Claim (1b) and (1c) basically follow directly from the definitions.

Claim (1d) is crucial for the construction of $\mathcal{A}'_\psi$, as it, intuitively, implies that, for each part of the data word, there are at most $\delta + 1$ data values that require special attention by $\mathcal{A}'_\psi$. To prove it we can assume that both $j$ and $j'$ have a non-empty herd as otherwise one of them has an interval of size at most $\delta$ and the claim follows trivially.

Let therefore $j < j'$ be two shepherds. Let $i = e^-(j)$, $k = e^+(j)$, $i' = e^-(j')$ and $k' = e^+(j')$. We prove $i' + \delta > k$ (and thus the claim) by contradiction. To this end, let us assume $i' + \delta \leq k < j < j'$.

| Pos  | $i$      | ... | $i'$       | ... | $i'+\delta$ | ... | $k$      | ... | $j$    | ... | $j'$   |
|------|----------|-----|------------|-----|-------------|-----|----------|-----|--------|-----|--------|
| Spec | $e^-(j)$ |     | $e^-(j')$  |     |             |     | $e^+(j)$ |     | $S$    |     | $S$    |
| Prop |          |     |            |     |             |     |          |     | $\psi$ |     | $\psi$ |
| $a$  | $d'$     |     | $d$        |     |             |     | $d'$     |     | $d$    |     |        |

Figure 3: Sketch for the proof of Claim (1d). Here, $S$ marks shepherds and $d, d'$ are two different data values.

As $j'$ is the $\psi$-shepherd for $i'$, $w[i'].@a = w[j].@a$ (as otherwise, because of $i' + \delta \leq j$, $j'$ would not be minimal). As $j$ is the $\psi$-shepherd for $i$, $w[i].@a \neq w[j].@a$. Thus, $w[k].@a = w[i].@a \neq w[j].@a = w[i'].@a$ and therefore $i' + \delta < k$. See Figure 3 for a sketch of the information obtained so far. However, if $k$ is $\rho$-special then for some $l \in [k, j) \subseteq [i + \delta, j)$ it holds $w[l].@a = w[k].@a$ and $w, l \models \rho^=$ but $w, l \not\models \rho^{\neq}$. As $w[l].@a = w[k].@a \neq w[i'].@a$, $j'$ cannot be a $\psi$-shepherd for $i'$, a contradiction. If, on the other hand, $k$ is $\tau$-special then there is a $\tau$-position $j'' \in [k, j)$ with $w[j''].@a = w[k].@a$. But then $j''$ would be a shepherd for $i'$, contradicting the minimality of $j'$. This concludes the proof of Claim (1d).

To deal with the additional complication added by $\delta$, we use additional propositions. In the intended string, each $\tau$-position carries a proposition $\tau_{(s)}$, for exactly one $s \in \{0, \ldots, \delta\}$. The herd of a shepherd marked by $\tau_{(s)}$ is called $s$-herd and so on. Besides $\tau_{(s)}$ we use further propositions of the form $\psi_{(s)}$, $e^+_{(s)}$ and $e^-_{(s)}$ with the intention that for each shepherd marked by $\tau_{(s)}$, $e^-(j)$ is marked by $e^-_{(s)}$ and $e^+(j)$ by $e^+_{(s)}$ and all positions in $H(j)$ are marked by $\psi_{(s)}$. We assume that all these propositions are already present in $w'$. However, the automaton has to test that they are as intended.

For a valid string $w'$ the assignment of $s$-values to shepherds can be done as follows. We assign numbers $s \in \{0, \ldots, \delta\}$ in a round-robin fashion to all positions of the form $e^+(j)$ and assign this $s$ to the corresponding $\tau$-position and its herd. This guarantees that for two positions $k = e^+_{(s)}(j)$ and $k' = e^+_{(s)}(j')$ with $j \neq j'$ we have $|k' - k| > \delta$. This implies that the intervals of two distinct $\tau_{(s)}$-positions are disjoint (as can be shown just the same way as Claim (1d)).

In a nutshell, the idea for the construction of $\mathcal{A}'_\psi$ is as follows. For each $s \in \{0, \ldots, \delta\}$ we construct an automaton $\mathcal{A}_{\psi,(s)}$ that takes care of $s$-shepherds and their herds. These automata independently check that for each $\tau_{(s)}$-position $j$ the corresponding $e^-(j)$- and $e^+(j)$-positions are correct, and guess and check all other positions in $S(j)$. Given the intervals, $\mathcal{A}_{\psi,(s)}$ checks that the $\psi_{(s)}$- and $\tau_{(s)}$-markings in $w'$ are consistent. The construction requires some more details to control inequality of data values sufficiently. By taking a suitable product of these $\delta + 1$ automata and additional auxiliary automaton (that tests, e.g., that the $e^+_{(s)}$-positions are numbered correctly) we get the automaton $\mathcal{A}'_\psi$.

In the following we describe the construction of $\mathcal{A}_{\psi,(s)}$ in more detail. We first define, for each attributed word $w'$ that is valid with respect to $\psi$ an extension $w''$ with further propositions. Then we formulate some conditions that are fulfilled by such a word $w''$ and show how they can be checked by a data automaton. Finally, we show that from a word $w''$ that fulfills the conditions a word that is valid with respect to $\psi$ can be extracted.

We use the additional propositions $\vdash, \bullet, \dashv$ to mark special positions in $s$-intervals and propositions $a^\rightarrow, a^\leftarrow, c^\rightarrow, c^\leftarrow$ to enforce some data inequality conditions. More precisely, the propositions $c^\rightarrow$ and $c^\leftarrow$ will help to make sure that every $s$-shepherd has a different data value from his herd and the propositions $a^\rightarrow$ and $a^\leftarrow$ will be used to pinpoint the $e^-_{(s)}$ positions.

To help the reader to distinguish the various kinds of propositions we use the following conventions.

- We say that positions are *labeled* by the propositions $\vdash, \bullet, \dashv$. We call a position that is not labeled by $\vdash, \bullet$ or $\dashv$ *unlabeled*.

- We say that positions are *colored* by the propositions $a^\rightarrow, a^\leftarrow, c^\rightarrow, c^\leftarrow$.

- For all other propositions we say that a position is *marked*.

In the following the *class $\delta$-predecessor* of a position $j$ is the largest position $k \leq j - \delta$ with $w'[k].@a = w'[j].@a$. We call the class 1-predecessor *class predecessor*

If $w'$ is a valid word (where the propositions $\tau_{(s)}, \psi_{(s)}, e^-_{(s)}, e^+_{(s)}$ are assigned as intended), we define its extension $w''$ as follows.

- For every $s$-shepherd $j$, $e^-(j)$ is labeled by $\vdash$ and $e^+(j)$ is labeled by $\dashv$. All special positions between $e^-(j)$ and $e^+(j)$ are labeled by $\bullet$.

- Let $j$ be the minimal $\tau_{(s)}$-position. The positions in $H(j)$ are colored by $c^\rightarrow$ and $j$ by $c^\leftarrow$. Then the following procedure decides the $c$-coloring for the remaining $\tau_{(s)}$-positions and their herds. Let $j$ be the smallest $\tau_{(s)}$-position whose $c^\leftarrow$-value has not yet been decided. Position $j$ is colored by $c^\leftarrow$ and all positions in $H(j)$ are colored by $c^\rightarrow$ if and only if its class $\delta$-predecessor is *not* colored by $c^\rightarrow$ or does not exist. As the herds of different $s$-shepherds are disjoint this procedure is unambiguous.

- The $a$-coloring is defined similarly. The first $\vdash$-position $k$ is colored by $a^\leftarrow$. All positions strictly between $k$ and its class predecessor are colored by $a^\rightarrow$. Afterwards, we proceed as follows. The minimal $\vdash$-position $k$ that is not yet decided is colored by $a^\leftarrow$ unless its class predecessor is colored by $a^\rightarrow$. All positions strictly between $k$ and its class predecessor which are larger than the previous $\vdash$-position are colored by $a^\rightarrow$ if and only if $k$ was colored by $a^\leftarrow$. Clearly, this procedure is unambiguous as well.

Next, we describe the conditions that are tested by $\mathcal{A}_{\psi,(s)} = (\mathcal{B}, \mathcal{C})$. For notational simplicity, we take the freedom to assume that the additional propositions are already given in the input, that is, $\mathcal{A}_{\psi,(s)}$ reads an extended word $w''$ instead of $w'$. However, it is straightforward to translate this automaton into one in which the base automaton guesses additional propositions (by means of states) and sends all propositions, encoded as states to the class automaton.

The first two conditions refer to the $a$- and $c$-colorings.

(Col 1) *(Consistent coloring for $\tau$)* The class $\delta$-predecessor of a $\tau_{(s)}$-position $j$ is colored by $c^{\rightarrow}$ if and only if $j$ is not colored by $c^{\leftarrow}$.

(Col 2) *(Consistent coloring for $\vdash$)* The class predecessor of a $\vdash$-position $j$ is colored by $a^{\rightarrow}$ if and only if $j$ is not colored by $a^{\leftarrow}$.

We say that a position $i$ is $a$-consistent with a position $k > i$ if $i$ is $a^{\rightarrow}$-colored if and only if $k$ is $a^{\leftarrow}$-colored. Likewise for $c$-consistent.

The next conditions deal with the intervals $I(j)$.

(Spec 1) *(Global $\vdash \bullet \dashv$ pattern)* In the whole attributed word, the labeled positions respect the pattern $((\vdash \bullet^* \dashv) + \{\vdash, \dashv\})^*$. Here, a position matches $\{\vdash, \dashv\}$ if both $\vdash$ and $\dashv$ occur.

(Spec 2) *(Local $\vdash \bullet \dashv$ pattern)* In each class word, the labeled positions occur in contiguous blocks each of which respects the pattern $(\vdash \bullet^* \dashv) + \{\vdash, \dashv\}$.

(Spec 3) *(Correct $\dashv$-positions)* A position $k$ is labeled by $\dashv$ if and only if there is a $\tau_{(s)}$-position $j > k$ and for the minimal such position $j$

- all $m \in (k, j)$ are $\rho^{\neq}$-marked and
  - $k$ is not $\rho^{\neq}$-marked or
  - $k$ is $\tau_{(s)}$-marked and
    * $j$ is unlabeled or
    * $j$ is labeled and there is a $\vdash$-position in $(k, j)$.

In this case, we call $j$ the corresponding $\tau_{(s)}$-position for $k$ (and for all other labeled positions in the interval of $k$).

(Spec 4) *(Correct $\vdash$-positions)* A labeled position $k$ is $\vdash$-labeled if and only if for its corresponding $\dashv$-position $m$ and its corresponding $\tau_{(s)}$-position $j$ the following conditions hold.

(A) all labeled positions in $(k + \delta, m]$ are $\rho^{=}$-marked,

(B) all unlabeled positions in $(k + \delta, m)$ are $\rho^{\neq}$-marked,

(C) all positions in $(m, j) \cap [k + \delta, j)$ are $\rho^{\neq}$-marked, and

(D) one of the following conditions holds.

(Di) There is a position $i < k$ such that $i + \delta < j$, $i + \delta$ is an unlabeled $\tau$-position and all positions in $(i, k)$ are $a$-consistent with $k$.

(Dii) There is a position $i < k$ such that $i + \delta < j$, $i + \delta$ is not a $\rho^=$-position and all positions in $(i, k)$ are $a$-consistent with $k$.

(Diii) There is a position $i < k$ such that $i + \delta < j$, $i + \delta$ is unlabeled and not marked $\rho^{\neq}$ and all positions in $(i, k)$ are $a$-consistent with $k$.

(Div) None of (i)-(iii) holds and all position $< k$ are $a$-consistent with $k$.

Finally, the following two conditions shall guarantee the validity of $w''$ with respect to $\psi$.

(Log 1) *(Unlabeled $\psi_{(s)}$-positions)* An unlabeled position $i$ is marked by $\psi_{(s)}$ if and only if there is a $\tau_{(s)}$-position $j \geq i + \delta$ and for the minimal such position it holds that every $k \in [i + \delta, j)$ is marked by $\rho^{\neq}$. If so, $i$ is $c$-consistent with $j$.

(Log 2) *(Labeled $\psi_{(s)}$-positions)* A labeled position $i$ is marked by $\psi_{(s)}$ if and only if there is a $\tau_{(s)}$-position $j \geq i + \delta$ after $i$'s corresponding $\dashv$-position $l$ and, for the minimal such $j$, every $k \in [i + \delta, l]$ is marked by $\rho^{\neq}$ or is labeled and marked by $\rho^=$ and every $k \in [l, j) \cap [i + \delta, j)$ is marked by $\rho^{\neq}$. If so, $i$ is $c$-consistent with $j$.

Conditions (Col 1), (Col 2) and (Spec 2) express properties of the class words of $w''$ whereas the remaining conditions express properties of its word projection. Each of these properties can be easily expressed in first-order logic and can therefore be checked by the class automaton or the base automaton, respectively. To test (Col 1) it is important that from the propositions $=_r$ it can be easily inferred how many positions are in the class word between a position $j$ and its class $\delta$-predecessor. Remember that $=_r$ is holds at position $i$ if and only if $w[i].@a = w[i + r].@a$ (see proof of Theorem 1).

We claim that the attributed word $w''$ constructed above fulfills all Col-, Ex- and Log-conditions if $w'$ is valid with respect to $\psi_{(s)}$. For (Col 1) this is because every shepherd is marked by $c^{\leftarrow}$ exactly if its class $\delta$-predecessor is not colored by $c^{\rightarrow}$ or does not exist at all. Likewise for (Col 2) and $\vdash$-positions. (Spec 1) holds because the intervals $I(j)$ are disjoint. (Spec 2) follows from the way the $\vdash \bullet \dashv$ labelings are obtained from the intervals. (Spec 3) holds by the definition of $e^+(j)$. (Spec 4) holds because the $\vdash$-positions are exactly the positions $e^-(j)$ and by the construction of the $a^{\rightarrow}$-coloring.

That (Log 1) and (Log 2) hold follows from the validity of $w'$ with respect to $\psi_{(s)}$ and the way the $c^{\rightarrow}$-marking is chosen.

Before we prove correctness for the automaton defined along these lines, we show the following useful claim

**Claim 2** *Let $w''$ be a 1-attributed word that fulfills all Col-, Ex- and Log-conditions and which is valid with respect to all subformulas of $\psi$. Then for each $\tau_{(s)}$-position $j$ the following conditions hold.*

(2a) There is a ⊣-labeled position $k$ for which $j$ is the corresponding $\tau_{(s)}$-position if and only if $S(j) \neq \emptyset$. If so, there is also a corresponding ⊢-position $l$, $k = e^+(j)$ and $l = e^-(j)$.

(2b) $S(j)$ is exactly the set of labeled points between the ⊢ and ⊣-position corresponding to $j$.

The part of Claim (2a) concerning $e^+(j)$ follows directly from (Spec 3), (Col 1) and (Log 2). This implies in particular that all labeled positions in some interval have a different data value than their corresponding $\tau_{(s)}$-position.

To show that $l = e^-(j)$ we have to show that (1) $l \in S(j)$ and (2) there is no smaller $l' \in S(j)$. By (Spec 1,2) all corresponding labeled positions have the same data value. Therefore (Spec 4 A-C) guarantee that $l \in S(j)$. For the sake of a contradiction let us now assume that there is $l' \in S(j)$ with $l' < l$. As $w''[l'].@a = w''[l].@a$, $l'$ is not $a$-consistent with $l$. By (Spec 4 D) therefore one of the following statements holds, each leading to a contradiction.

(1) There is an position $i \in [l', l)$ such that $i + \delta < j$ and $i + \delta$ is an unlabeled $\tau$-position. But then $i + \delta$ is a $\psi$-shepherd for $l'$ contradicting the minimal choice of $j$.

(2) There is a position $i \in [l', l)$ such that $i + \delta < j$ and $i + \delta$ is not a $\rho^=$-position. This contradicts $l' \in S(j)$.

(3) There is a position $i \in [l', l)$ such that $i + \delta < j$ and $i + \delta$ is unlabeled and not marked $\rho^{\neq}$. This also contradicts $l' \in S(j)$.

Before we show that the automaton $\mathcal{A}'_\psi$ is indeed correct, we make precise what it means that a 1-attributed word $w''$ is valid with respect to $\psi_{(s)}$ for a formula

$$\psi = (@a \wedge \rho^=) \vee (\overline{@a} \wedge \rho^{\neq}) U^{\neq} (\overline{@a} \wedge \tau).$$

This is the case if for every position $i$ of $w''$ it holds that $i$ is marked by $\psi_{(s)}$ if and only if

$(*_s^i)$ there is a position $j \geq i + \delta$ such that

  (I') $j$ is marked by $\tau_{(s)}$,
  (II') $w''[j].@a \neq w''[i].@a$, and
  (III') for the minimal such $j$ every position $k \in [i + \delta, j)$ fulfills
    (a) $k$ is marked by $\rho^{\neq}$ or
    (b) $w''[k].@a = w''[i].@a$ and $k$ is marked by $\rho^=$.

We show now that if a 1-attributed word fulfills the Col-, Ex- and Log-conditions above then each of its positions $i$ is marked by $\psi_{(s)}$ if and only if $(*_s^i)$ holds.

We first show that all $\psi_{(s)}$-positions $i$ fulfill $(*_s^i)$.

We first consider the case that $i$ is unlabeled. Let $j$ be the corresponding position guaranteed by (Log 1). Clearly $i$ and $j$ fulfill (I') and (III'). We have to show that $w''[i].@a \neq w''[j].@a$. To this end, let $k \in [i, j - \delta]$ be maximal with $w''[k].@a = w''[i].@a$.

- If $k$ is unlabeled then $k$ fulfills the condition of (Log 1) and is thus also labeled by $\psi_{(s)}$. Furthermore, it is $c$-consistent for $j$ and therefore $w''[j].@a \neq w''[k].@a = w''[i].@a$.

- If $k$ is labeled, there must be a $\vdash$-position $l \in (i, k]$. As $w''[i].@a = w''[k].@a = w''[l].@a$, $i$ is not $a$-consistent with $l$. By (Spec 4D) it follows again the existence of an intermediate position as in (1)-(3) in the proof of Claim (2a) and as there we get the desired contradiction.

We next consider the case that $i$ is labeled. Let $j$ be the corresponding position guaranteed by (Log 2). Clearly $i$ and $j$ fulfill (I') and (III') (as all labeled positions in $[i + \delta, l]$ have the same data value as $i$). Again, we have to show that $w''[i].@a \neq w''[j].@a$. Let $l$ be the $\dashv$-position corresponding to $i$, thus $l \in [i, j)$.

- If $j$ is labeled (Spec 3) guarantees that there is a $\vdash$-position $m \in (l, j)$. As $j$ is the first $\tau_{(s)}$-position after $l$ there cannot be any other $s$-interval between $l$ and $m$. (Spec 5) and (Col 2) now guarantee $w''[l].@a \neq w''[m].@a$ and thus $w''[i].@a = w''[l].@a \neq w''[m].@a = w''[j].@a$.

- If $j$ is unlabeled then $w''[j].@a \neq w''[l].@a = w''[i].@a$ because $j$ is the corresponding $\tau_{(s)}$-position for $l$.

Finally, we show that all positions $i$ fulfilling $(*_s^i)$ are marked by $\psi_{(s)}$.

Again, we consider first the case that $i$ is unlabeled. In this case $i \notin S(j)$. Therefore all intermediate positions $k \in [i + \delta, j)$ fulfill $\rho^{\neq}$. Therefore (Log 1) implies that $i$ is marked by $\psi_{(s)}$.

If, on the other hand, $i$ is labeled, $i \in S(j)$. Let $l = e^+(j)$, i.e., the corresponding $\dashv$-position of $i$. As all labeled positions in $[i + \delta, l]$ have the same data value as $i$, these positions fulfill the corresponding statement in (Log 2). Let us assume there is a position $m \in (l, j) \cap [i + \delta, j)$ that is not $\rho^{\neq}$-marked and fulfills $w''[m].@a = w''[i].@a$. But then $m$ is a $\rho$-stair for $i$ and therefore $m \in S(j)$. Thus, (Log 2) ensures that $i$ is $\psi_{(s)}$-marked. This completes the proof. □

By a straightforward extension of the proof of Theorem 2 we get the following.

**Theorem 8** *Satisfiability for XD-LTL on finite attributed words is decidable.*

**Proof.** It suffices to show that the construction of the proof of Theorem 2 can be extended for formulas of the new kind. To this end, we extend $t$ as follows.

$$t\bigg(\big(\rho \vee (@a_i \wedge \rho^=) \vee (\overline{@a_i} \wedge \rho^{\neq})\big) \operatorname{U}_{@a_j}(\overline{@a_k} \wedge \tau)\bigg) :=$$
$$t_j\bigg(\big((t(\rho) \vee \neg\text{att}_i) \vee (@a \wedge R \wedge t(\rho^=)) \vee (\overline{@a} \wedge R \wedge t(\rho^{\neq}))\big) \operatorname{U}_{@a}(\overline{@a} \wedge R \wedge \text{att}_k \wedge t(\tau))\bigg)$$

□

## 6 Conclusion

We conclude by stating some questions that should be investigated further. We would be interested to understand the exact border of undecidability. At this point, it is not exactly clear which kinds of intermediate and target formulas can be allowed for $\operatorname{U}_{@a}^{\delta}$. It would also be interesting to compare our logics with other logics that can deal with values, particularly with guarded LTL-FO of [14]. Further investigations could try to identify fragments with more reasonable complexity and try to add more arithmetics to the data domain.